\documentstyle[epsfig,multicol,aps,prb]{revtex}

\renewcommand{\appendix}{%
 \setcounter{section}{0}%
 \setcounter{equation}{0}%
 \renewcommand{\thesection}{{APPENDIX} \Alph{section}}
 \renewcommand{\theequation}{\Alph{section}.\arabic{equation}}%
}

\begin{document}
\preprint{LA-UR-01-805}
\title{\bf Theory of Semi-flexible polymers}
\author{Shirish M. Chitanvis}
\address{
Theoretical Division, 
Los Alamos National Laboratory\\
Los Alamos, New Mexico\ \ 87545\\}

\date{\today}
\maketitle
\begin{abstract}
We have mapped the physics of a system of semi-flexible
inextensible polymers onto a complex Ginzburg-Landau field theory
using techniques of functional integration.
It is shown in the limit of low number density of monomers in a melt
of semi-flexible, inextensible polymers, kept apart  
by an excluded volume interaction, the radius of gyration scales as
$N^{\nu}$, where $N$ is the chain length, and $\nu = 1$, in contrast
to the value of $\nu\approx 0.6$ for flexible polymer melts. 
Using Renormalization Group arguments, we show that the system
exhibits an infra-red stable fixed point which can be identified as
the transition to the entangled state.
Experiments to test these calculations are suggested.
\end{abstract}


\begin{multicols}{2}
\newpage
Polymers in a biological setting occur with a variety of persistence lengths, ranging
from being much smaller than the overall length of the polymer, to being
comparable to the chain length\cite{kroy_frey}.
When the persistence length is much smaller than the chain length,
the polymer can be described as being flexible.
Generally however, the persistence length is not much smaller than
the chain length, and the theoretical treatments
of such semi-flexible polymers becomes complex\cite{frey2,hinner}, especially when one takes into 
account the fact that these polymers are inextensible, and can
interact with each other.

One purpose of this paper is to understand the physics of melts of
inextensible, semi-flexible polymers in the low number density limit,
by generalizing de Genne's theoretical approach\cite{deG0}.
In this paper, 
it will be shown that the Flory exponent $\nu$, which characterizes the
radius of gyration $R_g \sim N^{\nu}$, where $N$
is total number of monomers in an average chain,
is different for the case of interacting semi-flexible polymers, and
is given by $\nu = 1$.

Renormalization Group arguments will also be used to show that as the
number density of semi-flexible polymer segments increases in a melt,
the system actually begins to behave as if the chains were becoming
increasingly independent, analogous to Flory's theorem for melts of
flexible chains.
Furthermore, it can be shown that the system scales towards a
thermodynamic fixed point as the number density is increased further,
indicating a second order phase transition to the entangled state.  

Treatments of semi-flexible polymers can become quite sophisticated,
even when restricted to a single chain\cite{wilhelm_frey,thiru,kleinert}.
In fact the treatment of ensembles of interacting semi-flexible chains is an
open problem.
In this paper, the focus will be on semi-flexible polymers interacting
through hard-core repulsion between the monomers.
This interaction will be represented by an excluded volume term.
It turns out that the elegant path integral formalism for a single
semi-flexible chain 
put forth by Kleinert\cite{kleinert} is well-suited for generalization
to a collection of interacting polymers.
In this generalization, the physics of semi-flexible polymers gets mapped onto a
time-dependent complex Ginzburg-Landau complex $\vert \psi\vert^4$
field theory.
Time in the theory is merely a label for monomers along the chain.

Let us begin with the worm-like model of Kratky and Porod for a single
semi-flexible polymer: 

\begin{eqnarray}
\beta {\cal H}_0 &&= {\kappa \over 2} \int_0 ^N dn ~ \left({\partial
    \vec t(n) \over \partial n}\right)^2 \nonumber\\ 
\vec t(n) &&= {\partial \vec R(n) \over \partial n}
\label{k-p}
\end{eqnarray}
where $\vec R (n)$ is the location of the $n^{th}$ monomer in the
chain which is $N$ units long, 
$\vec t(n)$ is the tangent vector at $\vec R (n)$, 
$\beta = 1/kT$,
$k $ is Boltzmann's constant, $T$ is the
temperature, 
$\kappa = \ell_p/b^3$, and 
$\ell_p$ is the persistence length over which the tangent vectors remain correlated,
and $b$ is the monomer length.
Note that Eqn.(\ref{k-p}) has been written in a form that is
reminiscent of that for the flexible chain case, with the
replacement $\vec R \to \vec t$.

Inextensibility can be incorporated formally as done by Kleinert\cite{kleinert} by
introducing a correlation function via a path-integral representation.
It expresses the fact that in $(\vec t,n)$ space, segments $q$ units
apart on the chain are distributed in a Gaussian fashion 
(see Eqn.(\ref{diff}) below).
This is esentially the definition of the persistence length along a
semi-flexible chain:

\begin{equation}
{\cal G}_0(\vec t(q), \vec t(0)) \sim \int_{\vec t(0)}^{\vec t(q)}
{\cal D}^{d-1} \vec t(n)~\exp(-\beta {\cal H}_0) 
\label{inext}
\end{equation}

where $d=3$ refers to the dimension of the tangent vector,
and integration over the reduced ($d-1$) dimensions incorporates a 
length preserving constraint.
More specifically, the restriction of the integration to $d-1=2$
dimensions indicates the incorporation of a local inextensibility
constraint, viz., $\vec t(n)^2 = (dx/ds)^2+(dy/ds)^2+(dz/ds)^2 = constant$.
This restriction indicates that the integration is to be performed
over the surface of a sphere in tangent space.
This can lead to cumbersome calculations.
An approximation will now be introduced, based on physical
considerations, 
which greatly reduces the level of complexity of the subsequent
calculations.

One can imagine that if we have a collection of worm-like chains,
such that it costs energy to bend any one of them, and
their number density is quite low, it would be quite easy for them
to remain straight (their lowest-energy configuration), as the excluded volume
term would not be a very effective constraint on their configuration.
In other words, the curvature of any given chain for low number
densities is expected to be low.
As a result, it follows that the radius of the sphere on which the
integration is to be carried out is quite large, compared to the
monomer length.
The surface is in this sense effectively flat, so that the integration can be
carried out in a flat two-dimensional manifold.
Equation(\ref{inext}) can now be evaluated easily:

\begin{equation} 
{\cal G}_0(\vec t(q), \vec t(0)) \equiv \left({2 \pi q b'^2 \over
    3}\right)^{-1}\exp\left[-{3 (\vec t(q)-\vec t(0))^2 \over 2 q b'^2}\right]
\label{diff}
\end{equation}
where $b'^2 = 3 b^3/\ell_p$.
Now this expression for the propagator is seen to be analogous to the
propagator ${\cal G}_{QM}$ for the Schroedinger equation in two dimensions, viz.,

\begin{equation}
{\cal G}_{QM}(\vec r(\tau),\vec r(0)) \equiv \left({-2 \pi \hbar \tau
    \over \tilde m}\right)^{-1}\exp\left[{i \tilde m \left(\vec 
    r(\tau)-\vec r(0)\right)^2 \over 2 \hbar \tau}\right]
\label{qm}
\end{equation}
where $\tau$ is the time, $\vec r$ is the position vector, 
and $\tilde m$ is the particle mass.
The mapping between the two models is given by  
$\tilde m \equiv {3 \over b'^2}$, and $i \tau \to-q$.

Since one can use Feynman's formulation to connect the
quantum-mechanical propagator to a wave function, it is natural to ask 
whether, working in imaginary time, one can relate the
Green's function in Eqn.(\ref{inext}) to a function that plays
the role of a probability amplitude.
It is advantageous to do so, since it permits the formulation of a
non-trivial model of interacting semi-flexible polymers. 

Our formulation uses a standard result from field theory\cite{kaku} to
express Eqn.(\ref{inext}) 
in an alternative form through the use of an auxiliary complex field
$\psi(\vec t,n)$, and an associated functional ${\cal F}_0$:

\begin{eqnarray}
&&{\cal G}_0(\vec t(N), \vec t(0)) \sim \int {\cal D}^{2}
\psi~\psi^*(\vec t,N)~\psi(\vec t,0)~\exp(-\beta {\cal F}_0)
\nonumber\\ 
&&\beta {\cal F}_0 = \int dn \int d^{d-1} \vec t~ \psi^*(\vec t,n)
\left(\partial_n - {b'^2 \over 6} \nabla^2 \right) \psi(\vec t,n)
\label{func}
\end{eqnarray}
where 
$\partial_n \equiv {\partial/\partial n}$,
$b'^2 = {3 b^2 /\lambda}$, and 
$\lambda = {\ell_p /b}$.

Even at this non-interacting level the new formulation has an advantage over
the preceding one in that it is a compact way of representing many 
independent polymers, since one can easily derive a $2p$-point correlation
functions as a product of $p$ factors of ${\cal G}_0$ by starting with the
partition function:

\begin{equation}
{\cal Z}_0 =  \int {\cal D}^{2} \psi~\exp(-\beta {\cal F}_0)
\label{z0}
\end{equation}

Excluded volume effects, representing the simplest form of interaction
between semi-flexible polymers can be expressed by the following extension:

\begin{eqnarray}
&&\beta {\cal F}_0 \to \beta {\cal F} = \beta {\cal F}_0 + \beta
\Delta {\cal F} \nonumber\\ 
&&\beta \Delta {\cal F} = \int dn \int d^{d-1} \vec t \left[ \left({v
      \over 2}\right) \vert \psi(\vec t,n)\vert^4 \ 
                        - \mu \vert \psi(\vec t,n)\vert^2 \right]  
\label{interact}
\end{eqnarray}

where $v$ is the usual excluded volume interaction parameter and $\mu$
is a Lagrange multiplier 
used to conserve the  number of monomer segments in the system.
The model was written down in analogy with the quartic quantum
electrodynamic theory, where  
an interaction term local in space and time suffices to describe the
physics, as long as the coupling constant is chosen correctly.
It can be thought of as a pseudopotential model which is local in the
independent variables.

An Euler-Lagrange equation obtained by
extremizing $\beta {\cal F}$ with respect to $\psi^*$.
An equation of continuity can be obtained as in Quantum mechanics
from it. 
$\vert \psi\vert^2$ can now be interpreted as a probability density:

\begin{eqnarray}
\partial_n \vert \psi \vert^2 &&= -\vec \nabla \cdot \vec j + S
\nonumber\\ 
\vec j &&= \left({b'^2 \over 6 }\right) \left(\psi^* \vec \nabla \psi
  + c.c.\right) \nonumber\\ 
S &&=  \left({b'^2 \over 6} \right) \vert \vec \nabla \psi \vert^2 + 2
\mu  \vert \psi \vert^2 - v \vert \psi \vert^4 
\label{cont}
\end{eqnarray}

While this interpretation of  $\vert \psi\vert^2$ as the probability density is
obtained within a mean field approximation, it should not be expected to break down
when we consider perturbative fluctuations around the mean field.

The chemical potential can be estimated by considering the homogeneous solution
of the Euler-Lagrange equation, which leads to $\mu = c_0 v$, where $c_0$ is
the mean number density.

Now let us re-examine Eqn.(\ref{interact}).
The integrand of $\beta \Delta {\cal F}$ is seen to exhibit a double minimum
potential as a function of  $\vert \psi \vert^2$.
As $\mu \to 0$, or equivalently $c_0 \to 0$, it can be seen that the double minimum
vanishes.  Thus this $\vert \psi \vert^4$ theory is seen to be analogous 
to a Ginzburg-Landau theory used to describe the approach of a thermodynamic
system to a critical point.
Consequently, for sufficiently small number densities, a system of semi-flexible
polymers approaches a critical point, where fluctuations become important.
We shall account for them by appealing to Renormalization Group
arguments.

Our interest is in systems composed of long chains.
As done previously\cite{shirish0}, we
use $\partial_n \to -i \omega$, upon invoking a Fourier transform
with respect to $n$.
Further, as usual,
$-i \omega \equiv \vert p-q \vert^{-1}$, where $\vert p-q \vert$ is the
separation of units along the chain.
The truncated Hamiltonian that will now be considered (for $\mu\to 0^+$)
is:

\begin{eqnarray}
&&\beta {\cal F}_{truncated} = \nonumber\\
&&\int d^{d-1}\vec t~\psi^*(\vec t)(N^{-1} -\nabla^2 + 
(v/2) \vert \psi(\vec t) \vert^2 ) \psi(\vec t)
\label{trunc}
\end{eqnarray}
where $N$ is the chain length in the system.
This truncation allows us to focus on the behavior of semi-flexible
segments separated by $N$ monomers, e.g., the end-to-end distance.
Now the entire Wilson formalism for dealing with second order phase
transitions can be brought to bear on the problem, and it follows that 
the characteristic length scale in the system $\xi$ possesses the
following scaling form, since $d-1=2$:

\begin{equation}
\xi \sim N
\label{scale}
\end{equation}

From de Genne's work\cite{deG0} and our own\cite{shirish0}, $\xi$ can
be identified with the 
radius of gyration.
If the system was composed of freely extensible chains, then the
restriction to $d-1$ dimensions would be unnecessary, and we obtain
the result that $R_g\sim N^{-0.63...}$.
It may turn out to be useful to study extensions of this theory to
non-integer dimensions, by restriciting the requisite integration to
$d-\epsilon$, where $1>\epsilon > 0$ indicates varying degrees of
extensibility of the chains.

As the number density increases, one can visualize these worm-like
chains beginning to 
get entangled, while attempting to retain their low-energy linear conformation,
by weaving themselves into a three-dimensional pile of enmeshed
semi-rigid needles.
To describe this transition, it suffices to consider the lowest order 
vertex correction, which arises from the first order bubble diagram.
It is convenient in this calculation to scale all lengths in the theory by
$c_0^{-1/3}$, where $c_0$ is the average number concentration of
segments in the system:

\begin{eqnarray}
&&\alpha \equiv c_0 v \to \alpha_R = \alpha - \left({3\over
    8}\right) \alpha^2 \Gamma(\vec k,\omega) \nonumber\\
&&\Gamma(\vec k,\omega) = -i\int_0^{2 \pi} d\phi \int_0^{\infty}
d\lambda \int_0^{\infty} dp p (2 \pi)^2 \theta(\mu - p^2/2m) \nonumber\\
&&\times \exp -i \lambda \left(y + k^2/2m + p^2/m + (p k/m) \cos \phi \right)
\label{vertex1}
\end{eqnarray}
where
$m = {3 /b'^2}$,
$y = -i \omega - 2 \mu$, and
the required frequency integral has been performed by the method of
residues by closing the contour in the upper half-plane, and the
denominator, viz., 
$\left(y + k^2/2m + p^2/m + (p k/m) \cos \phi \right) $
has been exponentiated using Feynman's trick.
The relevant integrals can be performed simply, with the result that:

\begin{eqnarray}
&&\Gamma(\vec k, \omega) = \left({m \over 4 \pi}\right)~ \times \nonumber\\
&&\ln \left[
{2 m y + 4 m \mu + \sqrt{k^4 + 4 k^2 m y + (2 m y + 4 m \mu)^2}
  \over k^2 + 4 m y} \right]
\label{vertex2}
\end{eqnarray}

Notice that as $k \to \infty$, $\Gamma(\vec k, \omega) \to 0$.
This is as it should be, since $\Gamma$ is simply the lowest order
density-density structure factor.

In fact, had we done the required frequency integral in
Eqn.(\ref{vertex1}) by closing the contour in the lower half-plane, we 
would have obtained, after performing dimensional regularization,
$\Gamma(\vec k, \omega) = \left({m \over 4 \pi}\right) 
\left( \psi(1) - \ln[m(y + 3 k^2/(4m))] \right)$,
where $\psi(1)$ is a digamma function.
This form does not vanish for large wavenumbers, as a density-density
structure factor should.

It follows immediately from Eqn.(\ref{vertex1}) and
Eqn.(\ref{vertex2}) that as the number density of segments increases,
the effective coupling constant decreases.
Physically this means that the semi-flexible polymers begin to behave
as if they were increasingly independent.
This is a direct generalization of Flory's theorem for
flexible polymer melts.

With Eqn.(\ref{vertex2}) in hand, it is possible to define a
beta-function as is done in Renormalization Group theory, to study the 
scaling properties of the renormalized coupling constant $\alpha_R$.

In the long wave-length limit, the structure factor $\Gamma(k=0,\omega)$
diverges logarithmically as $y \to 0^+$.
Note also that $y=(-i\omega-2 \mu)\equiv (1/N- 2 \mu)$ plays
the role of an effective inverse chain length.
We would like to show that this is a signal for the onset
of entanglement.
For this purpose we rescale $\alpha_R$, and other
relevant quantities, for convenience:

\begin{eqnarray}
\hat \alpha_R &&= \alpha_R \left({3 m \over 32 \pi {\cal Y}}\right) \nonumber\\
{\cal Y} &&\equiv {1 \over \ln \left[ 1/2 + {\mu \over y} \right]}
  \nonumber\\
\hat \alpha &&= \alpha \left({3 m \over 32 \pi {\cal Y}}\right) 
\label{scale2}
\end{eqnarray}
where ${\cal Y}$ also plays the role of an inverse length scale.
It has been introduced in order to facilitate a formulation of a scaling 
equation reminiscent of Wilson's Renormalization Group arguments.
A scaling function $\beta(\hat \alpha_R)$ can now be defined:

\begin{equation}
\beta(\hat \alpha_R) \equiv {d \hat \alpha_R \over d \ln {\cal Y}} = -
\hat \alpha_R +  \hat \alpha_R^2
\label{scale3}
\end{equation}

$\hat \alpha_R = 1$ is an infra-red stable fixed
point of the transformation (in terms of ${\cal Y}$), and signifies a
continuous phase transition.
This second order phase transition is different from the conventional
one in that the structure factor diverges merely logarithmically.
To connect this transition with the onset of entanglement in
semi-flexible polymers, we note that the system of semi-flexible
polymers can be described by the following effective energy functional:

\begin{eqnarray}
&&\beta {\cal F}_{effective} \approx \nonumber\\
&&\int d^{d-1} \vec k d\omega
\left( - \mu c(\vec k,\omega) + {1 \over 2}  c^*(\vec k,\omega)
  \Gamma(\vec k, \omega)^{-1} c(\vec k,\omega) \right)
\label{eff1}
\end{eqnarray}

The partition function can be easily computed, since only Gaussian
path integrals are involved.
The Helmholtz free energy $A$ follows from this:

\begin{equation}
A = - k T \ln(Z) \approx -\left({1\over 2}\right) c_0 k T V N
\mu^2 \Gamma(k=0,\omega=i/N)
\label{mech}
\end{equation}
Using $P = -\left({\partial A \over \partial V }\right)_{N,T}$ and
$B = c_0 \left( {\partial P \over \partial c_0 }\right)$,
we find:

\begin{equation}
B  = \left({(c_0 v)^2 (c_0^{1/3} {\ell_p}) k T N \over 8 \pi b^3}
  \right) \ln\left[{1\over 2}+{\mu \over 1/N - 2 \mu}\right]
\label{B}
\end{equation}

It follows that the bulk modulus exhibits a sharp increase as
$1/N -2 \mu \to 0^+$, as it must near the onset of
entanglement.
Thus we can identify the phase transition located using a
Renormalization Group argument with the onset of 
entanglement in a system of semi-flexible polymers.
The approximations utilized here are insufficient to describe the
mechanical properties of the system beyond the onset of entanglement.
The data provided by Hinner\cite{hinner} shows evidence
of a sharp increase of the modulus as entanglement is approached,
in a solution of actin.
There appear fluctuations in the data near the
transition point.
This is qualitatively consistent with the (logarithmic) divergence of
the structure factor at the transition point, which implies large
fluctuations in the density-density correlation function.
The theoretical curvature is also consistent with experimental data.
Note that the number density dependence of the bulk modulus is $\sim
c_0^{7/3}$, similar to the concentration power law dependence quoted
by Hinner et al\cite{hinner}.
It is difficult to perform a finer comparison with these data, since
the experiments were performed for solutions of
actin, and the theory presented here applies to melts.
It should be possible to use melts of semi-flexible polymers, where
the rigidity could stem from steric hindrance, for example, to test
the predictions of this model.

Hinner et al\cite{hinner} indicate that in the regime of entanglement,
past the critical point discussed above, semi-flexible polymers appear
to exist in tube-like cages.
In this regime, away from the critical point just discussed,
mean-field theory provides an adequate description.
To investigate the structure of these tubes, let us consider the
Euler-Lagrange which follows from extremizing ${\cal F}$ with respect
to $\psi^*$,
written down in dimensionless form, in cylindrical geometry:

\begin{equation}
{1\over \rho} {d \over d\rho}\left(\rho {d \psi_s(\rho)\over
    d\rho}\right) + \left(1 - {s^2\over \rho^2}\right)
\psi_s(\rho)-\psi_s(\rho)^3 = 0
\label{tube1}
\end{equation}
where the radial variable $\rho \equiv \vec t(n)\cdot \hat \rho$,
with $\hat \rho$ being the unit vector in the radial direction, and $s$
is an integer since a functional form $\exp(i s \phi)$ was assumed to
capture the azimuthal dependence of the field. 
The unit of length used for scaling is:

\begin{equation}
a = {b^{3/2}\over \sqrt{2 c_0 v \ell_p}}
\label{tube2}
\end{equation}

For $\rho \to 0^+$, $\psi(\rho)\sim A \rho^s \equiv A (\vec t(n)\cdot \hat \rho)^s$.
Here $A$ is an amplitude that can be adjusted in numerical
calculations to match smoothly\cite{shirish0}, over a distance $\sim
a$, the asymptotic 
solution for $\rho \to \infty$, viz.,
$\psi(\rho) = \pm \sqrt{1 - s^2/(\vec t(n)\cdot \hat \rho)^2}$.

Physically, what this mean-field solution indicates is that due to the inherent
stiffness of the polymer, it cannot be wound up too tightly.
In other words, for $\vec t(n)\cdot \hat \rho \equiv \partial_n \vec
R(n) \cdot \hat \rho \to 0^+$, the probability $\vert \psi \vert^2$ 
of finding a polymer in that tightly wound position is vanishingly
small.
On the other hand, we do not have free polymers.
There is an excluded volume interaction term which does not allow
the semi-flexible polymers to unfurl in a linear fashion, but are
instead forced to stay coiled, due to the presence of other polymers,
over a distance $\sim a$.
This healing distance $a$ can be regarded as a tube radius.
Equation(\ref{tube2}) yields an analytic dependence of this tube
radius on the number concentration of segments, persistence length,
etc.
If $\ell_p$ is of the order of the length of polymer chain, then
the tube radius is proportional to the inverse square-root of the
chain length.
Since the number density of segments cannot exceed
$1/v$, it follows that the smallest value of $a$ attainable is 
$b^{3/2}/\sqrt{2 \ell_p}$.

We have used field-theoretic techniques to provide an 
insight into the physics of semi-flexible polymer melts.
It will be interesting to see how this theory can be extended to study 
polyelectrolytes.

This research was supported by the Department of Energy contract
W-7405-ENG-36, under the aegis of the Los Alamos National Laboratory
LDRD-DR polymer aging program.




\end{multicols}
\end{document}